# Nonthermal Atmospheric Plasma Reactors for Hydrogen Production from Low-Density Polyethylene

Benard Tabu[1], Kevin Akers[1], Peng Yu[2], Mammadbaghir Baghirzade[3], Eric Brack[4], Christopher Drew[4], J. Hunter Mack[1], Hsi-Wu Wong[5], and Juan Pablo Trelles[1]

[1] Department of Mechanical Engineering, University of Massachusetts Lowell, MA, United States of America
[2] Department of Chemical and Biological Engineering, Tufts University, MA USA
[3] Department of Aerospace Engineering and Engineering Mechanics, University of Texas at Austin, TX, United States of America
[4] US Army Combat Capabilities Development Command (DEVCOM) Soldier Center, Natick, MA
[5] Department of Chemical Engineering, University of Massachusetts Lowell, MA, United States of America

## Abstract

Hydrogen is largely produced via natural gas reforming or electrochemical water-splitting, leaving organic solid feedstocks under-utilized. Plasma technology powered by renewable electricity can lead to the sustainable upcycling of plastic waste and production of green hydrogen. In this work, low-temperature atmospheric pressure plasma reactors based on transferred arc (transarc) and gliding arc (glidarc) discharges are designed, built, and characterized to produce hydrogen from low-density polyethylene (LDPE) as a model plastic waste. Experimental results show that hydrogen production rate and efficiency increase monotonically with increasing voltage level in both reactors, with the maximum hydrogen production of 0.33 and 0.42 mmol/g LDPE for transarc and glidarc reactors, respectively. For the transarc reactor, smaller electrode-feedstock spacing favors greater hydrogen production, whereas, for the glidarc reactor, greater hydrogen production is obtained at intermediate flow rates. The hydrogen production from LDPE is comparable despite the markedly different modes of operation between the two reactors.

## 1. Introduction

The global energy demand is projected to increase by 56% by 2040, driven by population growth and industrialization, particularly in developing countries [1, 2]. Since fossil fuels are responsible for 80% of global energy demands [3], greenhouse gas emissions and their adverse impacts are also expected to increase unless the production and use of alternative energy sources, such as green hydrogen, are scaled-up [4]. Hydrogen not only has the highest energy density (120 MJ/kg) of all fuels [5], but it also does not produce $CO_2$, the main greenhouse gas, when reacted with oxygen [5–7]. Furthermore, hydrogen is one of the most abundant elements in the earth's crust [9]. However, hydrogen does not occur naturally; instead, it is embedded in water, hydrocarbons, and solid organic compounds such as biomass and plastics [7–9].

The increasing global production of plastics, which surpassed 360 million tons in 2018 [8, 9], has led to a dramatic increase in plastic waste, polluting the environment and interfering with ecosystems [11, 12]. Incineration, the dominant approach to deal with plastic waste, leads to $CO_2$ emissions and is prone to emit volatile organic compounds deleterious to human health [17]. Strategies to valorize plastic waste, such as recycling and particularly its utilization as a source of hydrogen, could have a primary role in dealing with the disposal of plastic waste.

Traditional routes to produce hydrogen from plastic waste are mainly divided between thermochemical and electrochemical methods. In thermochemical approaches, heat is supplied to plastic waste to attain high temperatures (typically -3000 $^0$C) that promote desired chemical conversion reactions [18]. This can either be done in the absence of oxygen via pyrolysis [16–18] or in the presence of a controlled amount of oxygen through gasification [13, 17]. In electrochemical methods, plastic waste is converted directly or indirectly by reduction-oxidation reactions within electrochemical cells [18, 19]. Since both the electrolytes and electrodes require replenishing [23], electrochemical methods are generally more expensive than thermochemical approaches [24]. Even though thermochemical processes are widely used in plastic waste treatment, these processes typically depict low rates of hydrogen production, limited selectivity [24, 25], and low energy efficiency due to energy spent in auxiliary functions, such as cooling of gas products. Methods for plastic waste treatment based on low temperature and atmospheric pressure operation, such as nonthermal (low-temperature) plasma processes, have the potential to be more viable than current approaches [26–28]. Moreover, if powered by renewable electricity (e.g., wind or solar photovoltaic power), plasma-based techniques would mitigate $CO_2$ emissions associated with plastic waste treatment.

Plasma, i.e., partially ionized gas constituted of free electrons and heavy species (ions, atoms, and molecules), generated at (near) atmospheric pressure conditions is broadly classified as either thermal or nonthermal [30]. In thermal plasma, electrons and heavy species are in thermal equilibrium and therefore depict the same temperature, usually ranging from 6 000 to over 20 000 K [31]. In contrast, in nonthermal plasma, the temperature of free electrons is high (typically 1 eV ~ 11 600 K or higher) compared to the heavy species temperature (e.g., a few hundred Celsius), resulting in a state of nonthermal equilibrium [25, 26].

Plasma-based approaches for plastic waste treatment generally do not require oxidizing agents given the high reactivity promoted by plasma species. Moreover, atmospheric pressure plasma processes often have compact footprints thanks to the high fluxes of reactive species [27, 28]. The application of thermal plasma to plastic waste treatment has been studied to a significant extent, even leading to the construction of pilot plants [23, 36]. The high energy density and high temperature of thermal plasma processes are desirable for applications such as thermal sprays, welding, plasma cutting, and solid waste treatment. However, for processes that require selective treatment of reactants with relatively low melting points, such as hydrogen production from plastics, high-temperature operations may be undesirable as they may lead to limited energy efficiency or complex installations [37].

Approaches based on nonthermal plasma potentially have greater energy efficiency and selectivity than thermal plasma processes [30–32]. Furthermore, nonthermal plasma processes operating at atmospheric pressure are highly desirable due to potentially lower capital and operating expenses (e.g., no need for vacuum systems) and compatibility with other unit operations [40]. Yao and collaborators [41] studied the hydrogenolysis of polyethylene to light hydrocarbons using an atmospheric pressure nonthermal plasma reactor based on dielectric barrier discharge (DBD) over solid catalysts and using hydrogen and argon as the working gases. They obtained over 95% selectivity of lower alkanes ($C_1$-$C_3$) and low fractions (< 5%) of unsaturated hydrocarbons. Furthermore, their results showed that introducing a catalyst (Pt/C or SAPO-34) significantly improved the energy efficiency but had minimal influence on product formation rate. Aminu *et al.* [39] used a two-stage pyrolysis/low-temperature plasma catalytic process, also based on DBD, to produce hydrogen and syngas (a mixture of mainly hydrogen and carbon monoxide) from plastic waste. They concluded that low-temperature plasma enhanced the total gas production and hydrogen yield compared to the catalysis-only process. Also, syngas selectivity was greatest at 1 minute of operation, after which it declined due to the predominance of pyrolysis reactions. Diaz-Silvarrey *et al.* [15] pyrolyzed high-density polyethylene using a nitrogen DBD. They observed a significant increase in syngas production at moderate temperatures, i.e., from 15 wt% to 44 wt% at 600 °C. Xiao and collaborators [26] recovered hydrogen and aromatics from polypropylene waste via plasma-catalytic pyrolysis and noted an increment in the gas products of 18 wt% with 4.19 mmol/g $H_2$ formed. Ahmed *et al.* [42] critically reviewed plasma-based approaches for decomposing hydrocarbons and suggested using nonthermal plasma as an alternative to conventional catalytic decomposition methods. Although promising results have been reported in the literature, the potential of nonthermal plasma for plastic waste valorization is largely unexplored.

This article focuses on hydrogen production from low-density polyethylene (LDPE) using nonthermal atmospheric plasma. In addition to hydrogen, other co-products such as methane, ethylene, ethyne, propane, and larger molecular hydrocarbons have been reported from the processing of similar organic polymeric feedstock, such as high-density polyethylene (HDPE) [15], polyethylene [28], and polypropylene [43]. The present study focuses on the design and characterization of the plasma reactors for the production of hydrogen from LDPE, and therefore hydrogen is treated as the main product. LDPE comprises long hydrocarbon chains with short branches, usually between 0.5 and 1 million carbon units [44]. It is widely used for packaging, thin-film coatings, pipes, and cable production and is a primary component of global plastic waste [44]. This research is envisioned as an initial step toward valorizing plastic waste via the direct use of renewable electricity at atmospheric pressure and low-temperature conditions and with minimal auxiliary reactants. Section 2 presents the design of two nonthermal atmospheric pressure plasma reactors to produce hydrogen from LDPE. The experimental characterization of the reactors, encompassing electrical, fluid flow, and chemical diagnostics, is shown in section 3. Section 4 discusses the performance of the two reactors in terms of hydrogen production rate, production efficiency, and their correlation with operational parameters. Concluding remarks are presented in section 5.

## 2. Nonthermal plasma reactors

### 2.1 Reactors design

Two reactors are designed, built, and characterized for hydrogen production from atmospheric nonthermal plasma. The reactors are based on transferred arc (transarc) and gliding arc (glidarc) electrical discharges, and present complementary operational characteristics. Schematics of the reactors are presented in **Fig. 1**.

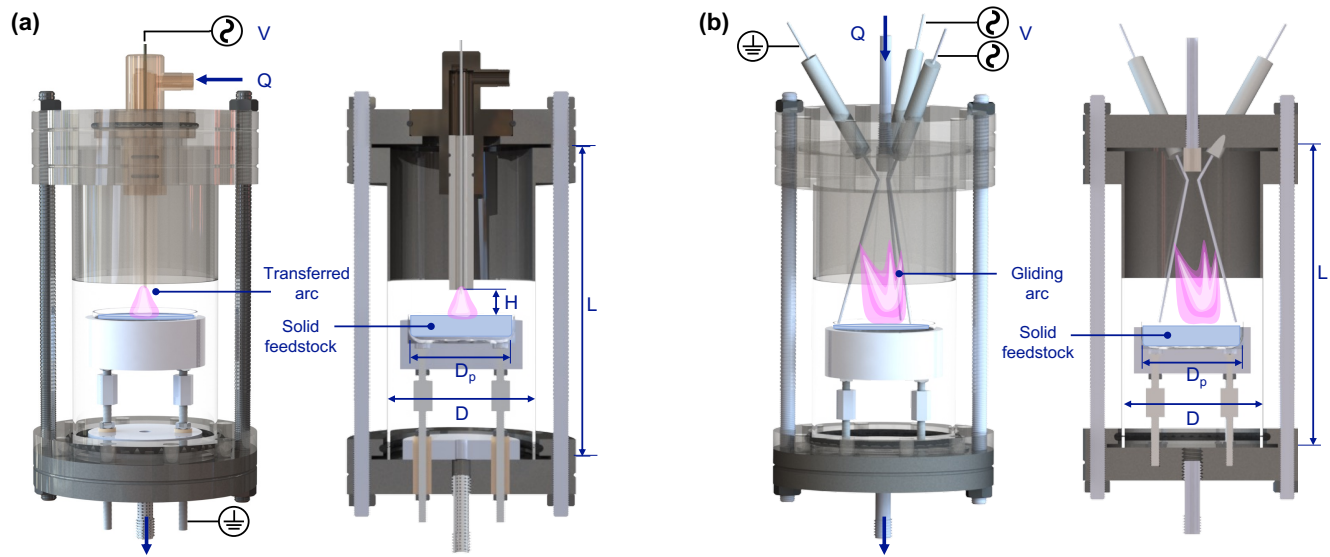

Fig. 1 - Plasma reactors for hydrogen production from polyethylene. Assembled design and cross-section view of the (a) transferred arc (transarc) reactor and the (b) gliding arc (glidarc) reactor. The reactors' main operating parameters are the electrode-feedstock spacing $H$, flow rate $Q$, and voltage level $V$.

The transarc reactor (**Fig. 1a**) has a pin-to-plate configuration with a powered tungsten electrode placed perpendicularly above an aluminum disc, which acts as the ground electrode and support of the crucible holding the solid feedstock (LDPE sample). The name *transferred arc* stems from electric current being *transferred* from the powered electrode to the feedstock. Thus, the feedstock is electrically coupled to the plasma. The distance between the tip of the powered electrode and the upper surface of the feedstock, denoted as *H*, is used as a control parameter. The gas nozzle is made of high-temperature resin fitted with a ceramic (alumina) bushing.

The glidarc reactor (**Fig. 1b**) consists of tri-prong equally spaced tungsten electrodes diverging 135° and with a gliding length of 30 mm. This electrode configuration generates a Y-shaped arc at the minimum inter-electrode separation distance. Two electrodes are powered by a separate power supply and the third electrode is set as ground. The name *gliding arc* stems from the fact that the generated arc *glides* along the electrodes due to the combined effects of advection of the gas inflow and the buoyancy of the low-density plasma. The minimum inter-electrode separation is 6 mm, as used in the glidarc reactor by Dassou *et al*. [45]. The glidarc plasma is electrically decoupled from the feedstock, making it suitable for treating a continuous stream of feedstock and surfaces.

Both reactors are powered by high voltage alternating current (AC) power supplies, delivering up to 300 W of output power with an independent frequency control from 20 to 70 kHz. The power supplies are voltage-controlled by setting the voltage level (*V*) from 0 to 100%, leading to a maximum voltage output (for zero load) from 1 to 40 kV. Nitrogen is used as a processing gas, injected with a flow rate (*Q*). The power supply voltage level *V* and flow rate *Q* are control parameters for both reactors. The reactors chambers have diameter *D* = 76.2 mm and height *L* = 150 mm and a quartz section to allow optical access. The residence time $t_{res}$ of the gas is therefore given by:

$$t_{res} = \frac{\pi D^2 L}{4Q}. \qquad (1)$$

The solid LDPE samples have a fixed mass of 10 g and are placed inside a quartz plate 55 mm in diameter and 15 mm in height. The quartz plate with the feedstock is fitted in a cylindrical aluminum holder. The holder acts as the ground electrode for the transarc reactor, but it is electrically de-coupled in the glidarc reactor.

Computational Fluid Dynamics (CFD) thermal-fluid models created in SolidWorks Flow Simulation [46] are used to evaluate the effect of control parameters on the operation of the reactors. The models describe the plasma as a volumetric heat source approximated as a solid with 100% porosity (i.e., no inertial resistance to fluid transport) in chemical equilibrium (i.e., species composition and material properties are a function of the local temperature only). For the transarc reactor, the plasma is approximated as a rectangular cylinder of 1.6 mm diameter connecting the tip of the powered electrode to the feedstock. Whereas for the glidarc reactor, the plasma volume is approximated as a truncated pyramid with a triangular cross-section 50 mm long approximately filling the inter-electrode space. Convective heat transfer boundary conditions, specified with an outside temperature of 300 K and a convective heat transfer coefficient of 25 W/m$^2$K, are imposed over all the outer surfaces of the reactors. Given the chemical equilibrium assumption, no chemical kinetics associated with the plasma or the interaction between the plasma and the feedstock are explicitly included in the models. Instead, the thermal-fluid models describe fluid flow and thermal characteristics throughout the reactors (reactor chamber, solid feedstock, and auxiliary components). Given the nonthermal nature of the generated plasma in the reactors, only a portion of consumed power is dissipated as heat. The amount of thermal power (dissipated heat) is an input to the models. Therefore, the models describe the operation of the reactors as a function of the control parameters inflow rate *Q* and thermal power dissipated by the plasma (assumed correlated with *V*). Representative results of the thermal-fluid models are presented in **Fig. 2**.

**Fig. 2** shows the geometry of the computational thermal-fluid models, as well as velocity and temperature distributions for representative operating conditions, namely dissipated thermal power of 1.75 W (i.e., 5% of 35 W, a representative value of power consumed by the plasma), nitrogen flow rate *Q* = 0.1 slpm, and electrode-feedstock spacing *H* = 5 mm for the transarc; and input power of 1.4 W (i.e., 5% of 28 W), nitrogen flow rate *Q* = 2 slpm, and electrode-feedstock spacing *H* = 5 mm for the transarc and glidarc. The transarc temperature distribution (**Fig. 2b**) is highest at the center of the plasma volume and decreases uniformly with increasing radial distance. This highest temperature on the feedstock surface is ~ 750 K. The relatively high temperatures in the transarc simulations are attributed to the relatively small plasma volume, which leads to increased thermal power per unit volume. The temperature distribution for the glidarc reactor presents a three-fold symmetry, which suggests non-uniform heating of the feedstock (**Fig. 2e**), with the highest temperature over the feedstock close to 300 K. The simulation predicts a significantly greater area of plasma interaction with the feedstock's surface as compared to the transarc reactor. This observation is complemented by the isosurface temperature distributions shown in **Fig. 2a** and **2d**.

Despite the high temperatures in the plasma volume, particularly for the transarc, the temperature near the reactors' walls is close to the ambient temperature of 300 K, irrespective of the amount of imposed thermal power. This suggests that the reactors can operate at or near room temperature without forced cooling. The flow fields in **Fig. 2c** and **2f** show that the axial gas inflow leads to the formation of vortex rings near the surface of the sample in both reactors. These vortical structures are characterized by relatively long residence times and may lead to the recombination of gas products emanating from the feedstock. Moreover, the higher velocity at the center of the transarc reactor indicates the potential formation of a crater-like pattern at the center of the feedstock. In contrast, the low-velocity magnitude of the three-fold way over the feedstock surface observed in the glidarc simulations suggests a more uniform treatment. These model predictions are contrasted against experimental observations in section 3.4 and section 4.1, respectively.

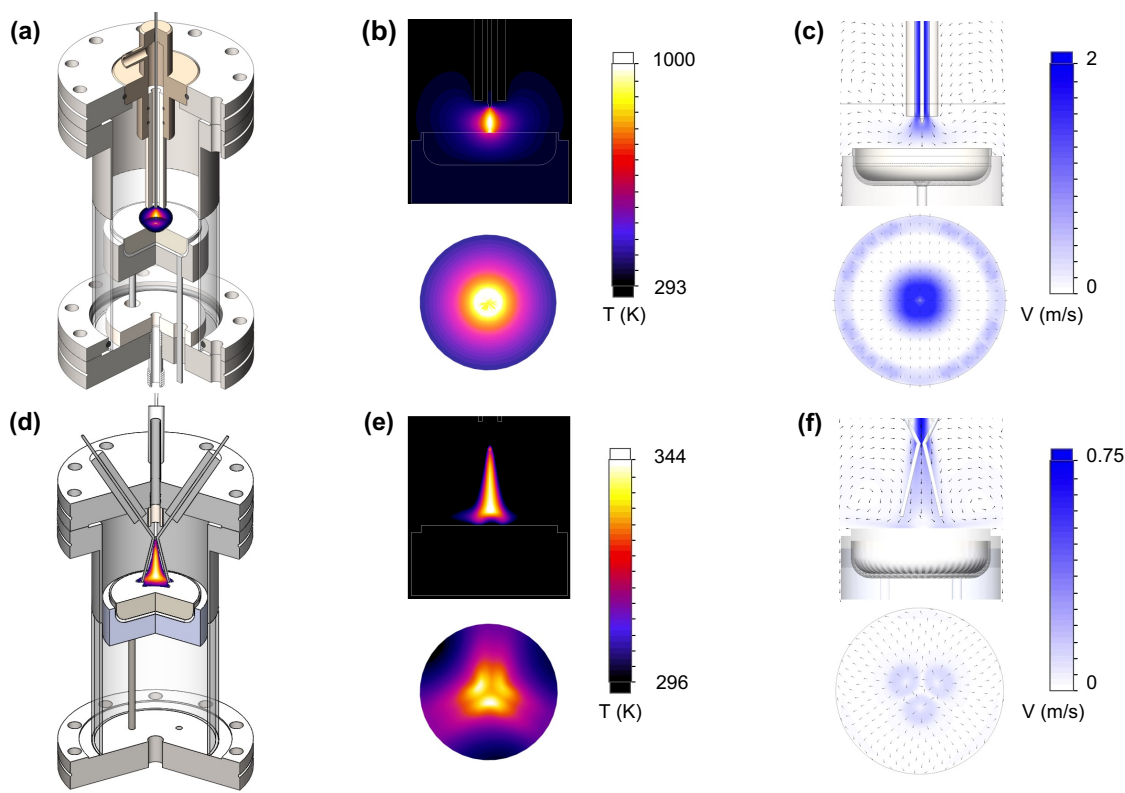

**Fig. 2 - Computational thermal-fluid reactor models. Transarc reactor: (a) design schematic, (b) temperature distribution, and (c) velocity distribution for 1.75 W of heat dissipation, $Q$ = 0.1 slpm, and $H$ = 5 mm. Glidarc reactor: (d) design schematic, (e) temperature distribution, and (f) velocity distribution for 1.4 W of heat dissipation, $Q$ = 2 slpm, and $H$ = 5 mm.**

## 2.2 Operational characteristics

The expected operational characteristics of the reactors obtained with the thermal-fluid models as function of dissipated thermal power and flow rate are shown in **Fig. 3**. For the transarc reactor, the average surface temperature increases linearly with dissipated thermal power per unit volume (**Fig. 3a**) and has minimal dependence on flow rate. The small difference in the average surface temperature for the flow rate of 2 and 4 slpm at $0.8\times10^3$ W/cm$^3$ is ascribed to the computational error. The glidarc reactor's average surface temperature varies directly with dissipated thermal power, but inversely with flow rate, as shown in **Fig. 3b**. A higher flow rate leads to enhanced convective cooling, which reduces the amount of heat deposited on the substrate. The average heat flux over the feedstock for both reactors (**Fig. 3c** and **Fig. 3d**) follows the same trends as the average surface temperature. These simulation results suggest that the transarc reactor can operate with small flow rates compared to those needed for the glidarc reactor, whose operation is very sensitive to flow rate. Based on these results, the experimental characterization of the reactors uses voltage level *V* (assumed proportional to thermal power dissipation) for both reactors, electrode-feedstock spacing *H* for the transarc reactor, and flow rate *Q* for the glidarc reactor, as main operational parameters.

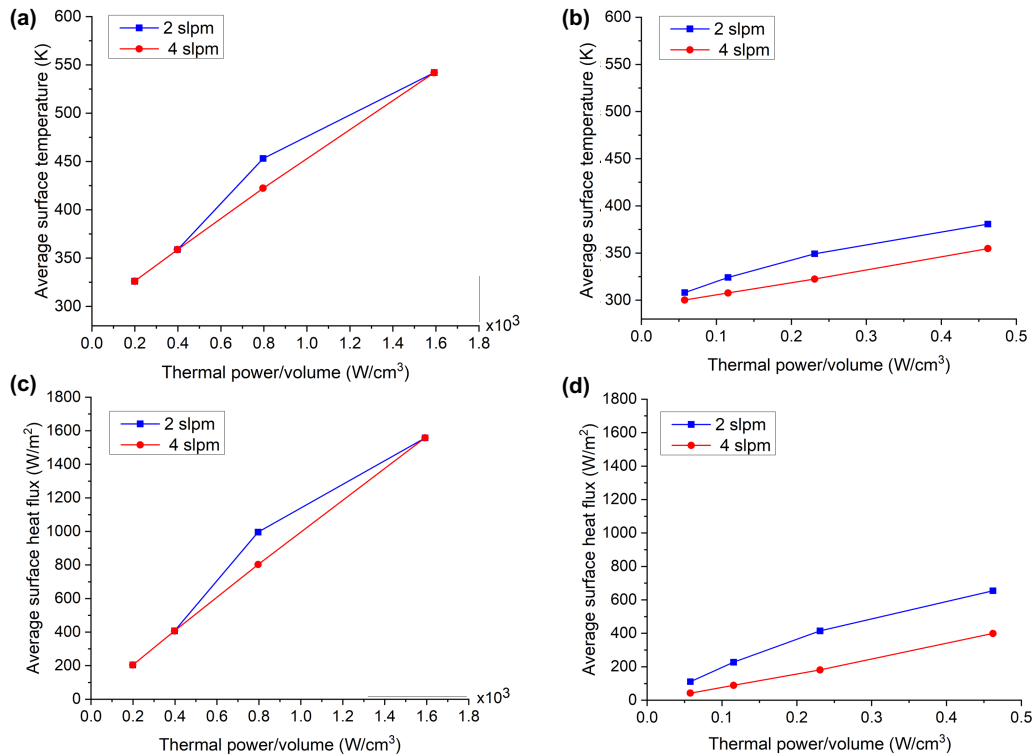

**Fig. 3 - Operational characteristics predicted by thermal-fluid models. Average surface temperature for the (a) transarc and (b) glidarc reactors and average surface heat flux for the (c) transarc and (d) glidarc reactors for varying thermal power density (proportional to *V*) and flow rate *Q*.**

## 3. Characterization of reactors

### 3.1 Experimental set-up

The experimental set-ups are depicted in **Fig. 4**, one for characterizing the operation of the reactors (**Fig. 4a**) and the other for Schlieren imaging (**Fig. 4b**). The reactors are powered by high voltage AC power supplies (PVM500-2500 Plasma Power Generator) with peak-to-peak voltage from 1 to 40 kV, and 25 mA peak current. As indicated in section 2, the transarc reactor operates with a single power supply, whereas the glidarc reactor utilizes two power supplies. Two Alicat mass flow controllers regulate the gas flow rate through the reactor; one is used for the nitrogen inflow and the other for the gas products outflow. A Tektronix Oscilloscope (TBS 2104) equipped with a current probe (P6021A) and high voltage probe (P6015A) are used to measure the electrical characteristics of the reactors' operation. The gas products are analyzed by a Shimadzu GC-2014 Gas Chromatography (GC) system.

The Schlieren imaging set-up allows the visualization of refractive index variations, which depict density gradients in the test medium. The set-up consists of collimator and de-collimator lenses (with focal lengths of 30 and 50 cm, respectively) aligned with a light fiber-optic and halogen source (250 W), the test medium (center of the plasma region within the reactor chamber), a knife-edge, and a high-speed camera (Edgertronic SC2+). The knife-edge adjusts the system's sensitivity while the high-speed camera captures the density gradient variation of the test medium. The test medium comprises the plasma interacting with either feedstock (LDPE) or an inert (quartz disc) sample within cross-shaped reactor chambers with flat quartz windows (to prevent optical distortions by the curvature of cylindrical quartz chambers). The high-speed camera is configured with shutter speed and frame rate of 1/8500 s and 8000 fps, respectively, to visualize the transarc plasma. To visualize the glidarc plasma, due to its dynamic nature with a gliding period in the order of milliseconds, a lower frame rate of 500 fps is used. The optical visualization of the operation of both reactors uses a camera adjusted to 1080 p resolution and 30 fps.

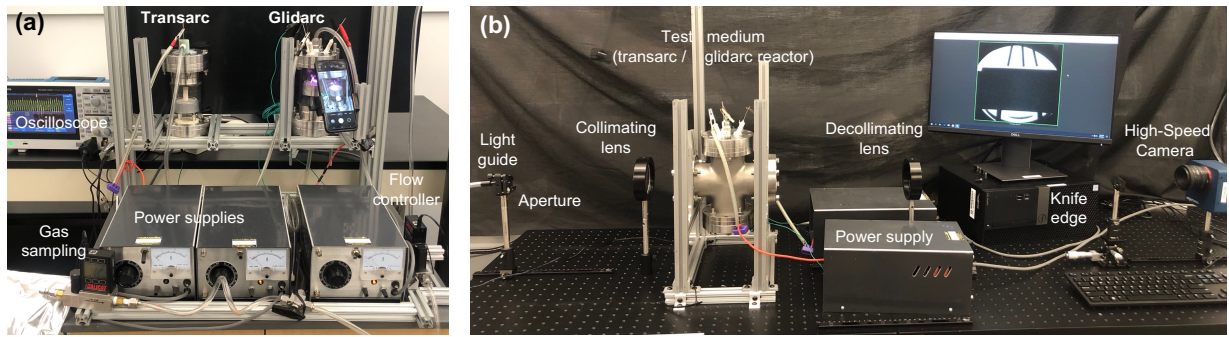

**Fig. 4 - Experimental set-ups for (a) characterization of the operation of the reactors during hydrogen production and for (b) optical and Schlieren visualization.**

### 3.2 Optical imaging

Optical characterization of the operation of the transarc reactor is conducted for varying voltage level $V$ and flow rate $Q$, as depicted in **Fig. 5**. The electrode-feedstock spacing $H$ is kept fixed at 10 mm. The minimum and the maximum $V$ are first determined for each $Q$. Given that the computational characterization of the transarc reactor showed a limited effect on the flow rate (**Fig. 3**), three relatively small flow rates, i.e., 0.1, 0.5, and 1.0 slpm, are used. These flow rates correspond to residence times $t_{res}$ of approximately 410, 82, and 41 s, respectively. The minimum and maximum voltage levels are set equal to 6 and 30%, respectively, for all flow rates. The minimum voltage level at a given $Q$ leads to faintly visible plasma (i.e., corona discharge). The intensity of the discharge increases with voltage level leading to transition from corona ($V$ = 6% to 10%) to glow ($V$ = 10% to 20%), and then to arc/spark discharge ($V$ > 20%). Discernably, a flow rate of 1 slpm produces a less-intense discharge with a slightly larger divergence of the plasma column than the discharges at 0.1 and 0.5 slpm. The intensity and divergence of the discharge are identified as critical parameters for hydrogen production. Hence, based on the optical characterization results, a fixed value of $Q$ = 0.1 slpm and $V$ = 20% and 30% are chosen for the hydrogen production experiments. To expand the range of characterization of the transarc reactor operation, given its minor sensitivity to $Q$, the electrode-feedstock spacing $H$ is set to either 5 or 10 mm in the hydrogen production experiments.

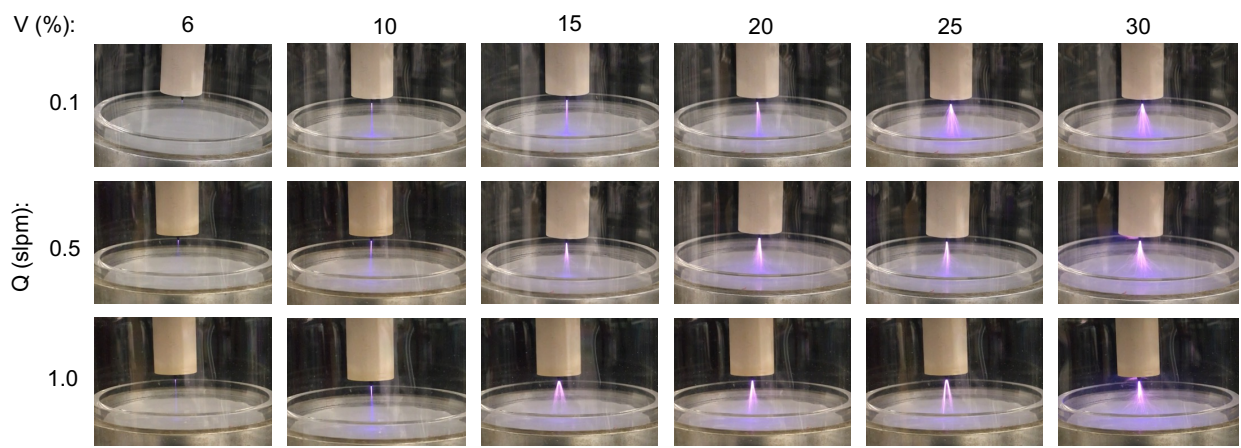

**Fig. 5** - Characterization of the transarc reactor. Optical imaging for varying flow rate $Q$ and voltage level $V$ for $H = 10$ mm, depicting a sparky discharge in all the operating conditions.

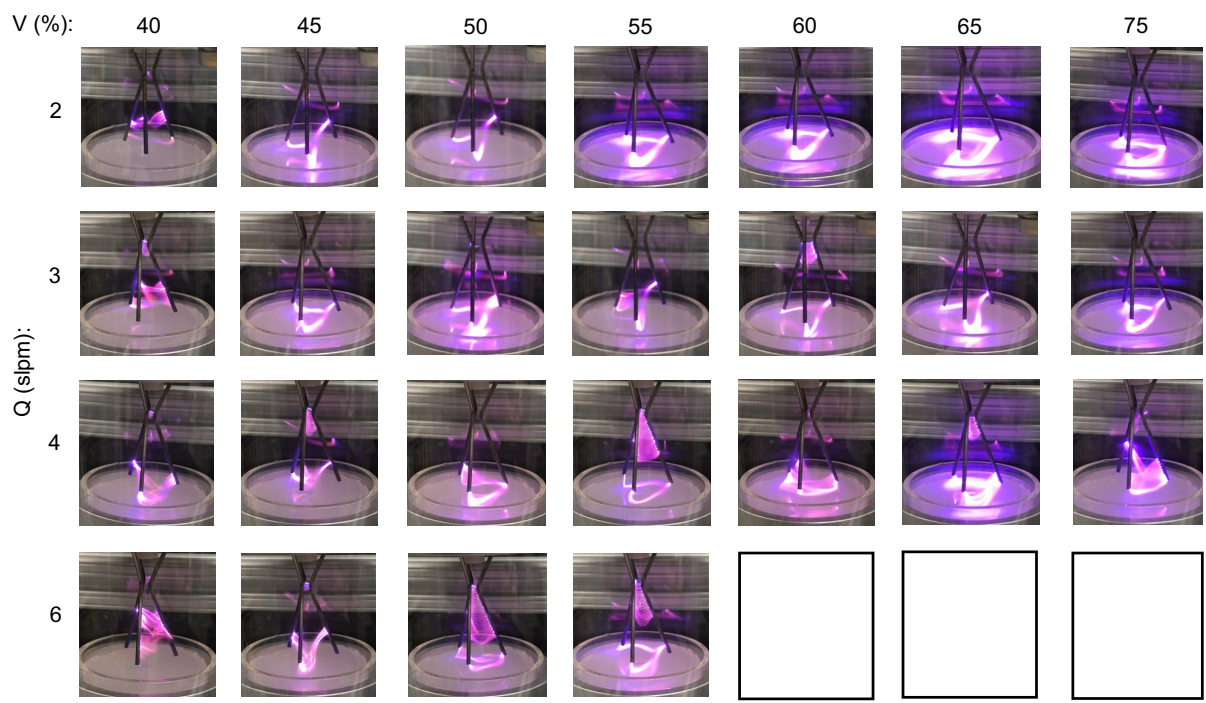

**Fig. 6** - Characterization of the glidarc reactor. Optical visualization for varying flow rate $Q$ and voltage level $V$. (No plasma is generated for $Q = 6$ slpm and $V = 60\%$ and higher).

In contrast to the transarc reactor, the glidarc reactor requires higher flow rates to establish appropriate interaction between the plasma and feedstock. Therefore, larger $Q$ values, i.e., 2, 3, 4, and 6 slpm, are used in the experimental characterization. These flow rates correspond to residence times $t_{res}$ of 20.5, 13.7, 10.3, and 6.8 s, respectively. The lowest and highest $V$ is 40% and 75%, respectively, across all flow rates. The results of the characterization of the operation of the glidarc reactor by optical imaging are shown in **Fig. 6**. The intensity of the tri-prong arc proportionally increases with voltage level for all investigated flow rates. The plasma does not interact with the feedstock for the highest flow rate of $Q = 6$ slpm and $V \geq 60\%$ or flow rates $Q < 2$ slpm at any $V$. This behavior is a characteristic of gliding arc discharges, whose dynamics depend on the balance between buoyancy and advective forces (due to the low density of the plasma and due to the drag by the gas flow, respectively). For $Q$ < 6 slpm, the impingement of the plasma on the feedstock is more pronounced for $V \geq 60\%$, suggesting that greater $V$ would favor greater hydrogen production. Therefore, for the hydrogen production tests, the glidarc reactor is operated at higher voltage levels of $V = 65\%$ and $75\%$, and flow rates $Q = 2$ and $4$ slpm.

**3.3 Plasma-feedstock interaction**

The solid feedstock used in the hydrogen production experiments consists of commercial LDPE pellets (average diameter of 3 mm) from Millipore Sigma (Sigma Aldrich, 428043). LDPE pellets totaling 10 g are melted at 180 $^0$C in the quartz plate ($D_p = 55$ mm diameter, see **Fig. 2**) using an electrical heater (Fisherbrand, 100-120 V) and then allowed to solidify to yield a solid LDPE sample of approximately 6 mm thickness.

The hydrogen production experiments consist of treating the solid LDPE sample with nitrogen plasma in either the transarc or glidarc reactor for 30 minutes. Gas product samples are extracted at 5-minute intervals throughout the experiments. The experiment for each set of operating conditions (i.e., $V$ and $H$ for the transarc and $V$ and $Q$ for the glidarc) is repeated three times. The variation in results is quantified by the error bars (i.e., standard error of the mean). In the hydrogen production experiments, the transarc reactor is operated under a low flow rate of 0.1 slpm while varying voltage level ($V = 20\%$ or $30\%$) and electrode-feedstock spacing ($H = 5$ or $10$ mm). For the glidarc reactor, a fixed electrode-feedstock spacing of 5 mm is used with varying flow rate ($Q = 2$ or $4$ slpm) and voltage level ($V = 65\%$ or $75\%$). These conditions are selected based on results in section 2 and section 3.3.

Representative images of the operation of the reactors at the beginning (0.5 min) and the end (30 min) of the hydrogen production experiments are shown in **Fig. 7.** Optical images of the transarc at 0.5 and 30 minutes of operation are depicted in **Fig. 7a** and **7b**. The results show that for $H = 5$ mm, the plasma presents a stable and intense glow, whereas for $H = 10$ mm, the plasma appears filamentary, representative of arc/spark conditions. This filamentary arc covers a broader area of the sample's surface, potentially leading to greater hydrogen production (section 4). The LDPE sample melted after ~ 5 minutes of operation. The yellow glow by the end of the experiment for $H = 5$ mm and $V = 30\%$ (**Fig. 7b**) can be attributed to the emission from carbon particles. For the larger spacing of 10 mm, a filamentary discharge weakly impinges the surface of the LDPE, and the plasma characteristics (size, emission, and dynamics) minimally change during the duration of the experiments.

The glidarc reactor generates a tri-prong arc that impinges on the feedstock at the end of each gliding period, as shown in **Fig. 7c** and **7d**. The intensity of the glidarc plasma is higher for $Q = 2$ slpm, which is credited to the longer residence time, leading to pronounced interaction with the LDPE sample and, consequently, higher hydrogen production (section 4). The experiments show significant differences between the plasma near the initial and final portions of the LDPE treatment. This is attributed to the formation of gaseous products, heating of the feedstock, and heating of electrodes. The blue-green glow over the sample's surface for $Q = 2$ slpm and $V = 75\%$, which is not observed under any other operational condition, suggests the formation of hydrocarbons.

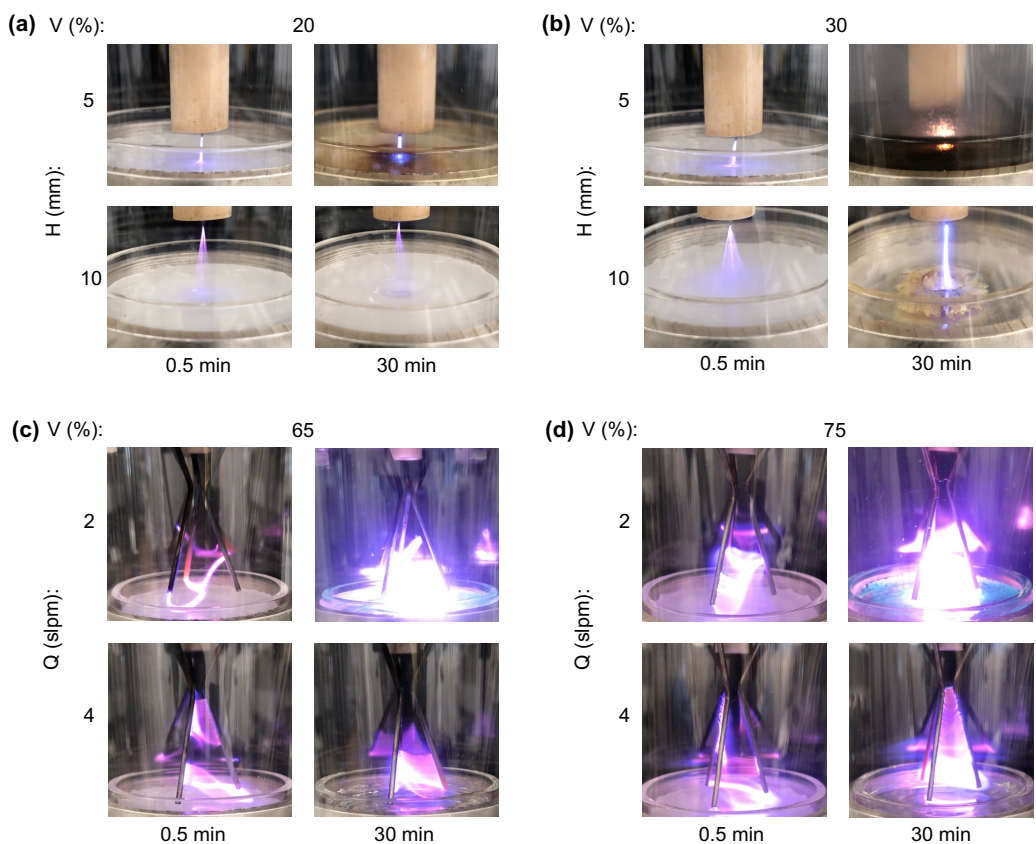

**Fig. 7** - Reactors operation during hydrogen production from LDPE. Transarc reactor at (a) the beginning and (b) the end of the experiment. Glidarc reactor at (c) the beginning and (d) the end of the experiment.

### 3.4 Schlieren imaging

Schlieren imaging allows resolving the flow dynamics inside the reactor and unveils potential relationships between plasma dynamics and hydrogen production. Schlieren imaging results of the transarc reactor interacting with the inert (quartz) and LDPE samples are presented in **Fig. 8a** and **Fig. 8b**, respectively. The transarc plasma interacting with the inert and LDPE samples, as indicated by the horizontal (green) arrow, is faintly visible for every experimental condition. The interaction of the transarc plasma with the inert sample generates mild turbulence, which is weakly visible in **Fig. 8a**. However, the transarc plasma interaction with the LDPE sample (**Fig. 8b**) produces significant turbulence over the surface of the feedstock, as indicated by the vertical (purple) arrow. Given that Schlieren imaging resolves mass density gradients within the flow and that hydrogen is significantly lighter than nitrogen (the working gas), the observed turbulence is probably due to hydrogen emanating from the surface of the feedstock. The more significant turbulence observed for the condition of $H = 5$ mm and $V = 30\%$ is consistent with greater hydrogen production (discussed in section 4.2).

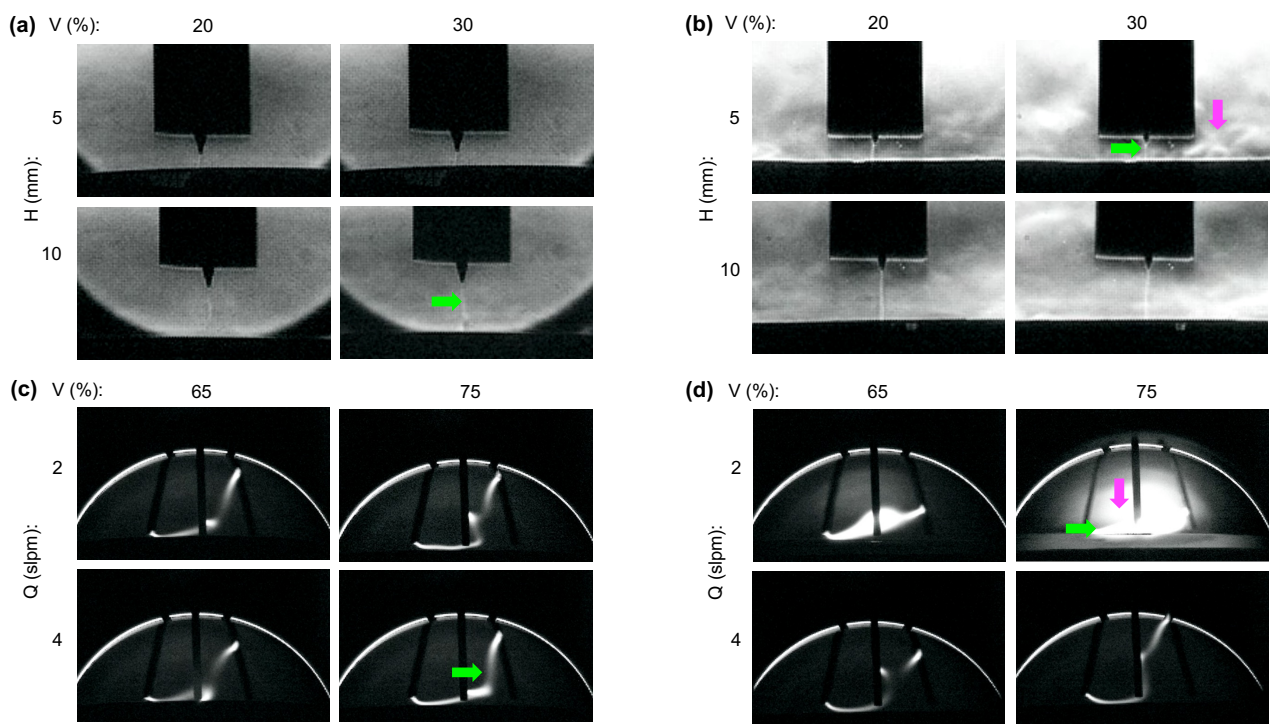

**Fig. 8** - Schlieren imaging of the operation of the reactors. The transarc plasma interacting with (a) inert and (b) LDPE samples. Glidarc plasma interacting with (c) inert and (d) LDPE samples. The horizontal arrows indicate the location of the plasma column, and the vertical arrows indicate the formation of turbulent flow originating from the surface of the sample, likely due to the production of hydrogen.

Schlieren imaging results of the glidarc reactor reveal the gliding of the tri-prongs arc (indicated by the horizontal green arrow) and its eventual impingement onto the sample, as shown in **Fig. 8c** and **Fig. 8d** for the inert and LDPE samples, respectively. The occurrence of turbulence is not captured by Schlieren imaging due to the significantly slower dynamics of the glidarc than those for the transarc (i.e., a 16 times lower frame rate is used to capture the dynamics of the glidarc than that used for the transarc). The interaction of the glidarc plasma with the inert sample (**Fig. 8c**) does not generate any glow, and the arc extinguishes on reaching the sample's surface. This suggests that the inert sample does not produce a significant amount of hydrogen when interacting with plasma. In contrast, when the glidarc plasma interacts with the LDPE sample, a pronounced glow is observed, indicated by the vertical (purple) arrow shown in **Fig. 8d**. The pronounced glow observed for $Q = 2$ slpm is probably attributed to hydrogen emanating from the feedstock, which correlates with greater hydrogen production (section 4.2). The horizontal arrows indicate the location of the plasma column, and the vertical arrows show the formation of turbulent flow from the substrate, likely due to the production of hydrogen.

## 4. Hydrogen production from LDPE

### 4.1 Sample characterization

The treated samples depict the extent of interaction between the plasma and the LDPE feedstock. **Fig. 9** shows the LDPE samples before (**Fig. 9a**) and after 30 minutes of treatment (**Fig. 9b** for the transarc and **Fig. 9c** for the glidarc) for the selected values of operational parameters. For the transarc, the white surface of the pristine LDPE sample develops a dark-brown color after treatment, especially for $H$ = 5 mm (**Fig. 9b**). The significantly darker and more extensive region of the sample treated using $H$ = 5 mm and $V$ = 30% implies a more significant plasma-LDPE interaction, consistent with the observed greater turbulence (**Fig. 8b**). The dark color suggests the formation of carbon compounds over the treated feedstock surface, consistent with the emission of carbon particles implied by the results in **Fig. 7b**. For $H$ = 10 mm, the weak spark discharge generated leads to the formation of a crater at the center of the sample. This crater formation is suggested by the computational simulation results in section 2.2, which show concentrated temperature, heat flux, and velocity at the center of the feedstock (e.g., **Fig. 2b**).

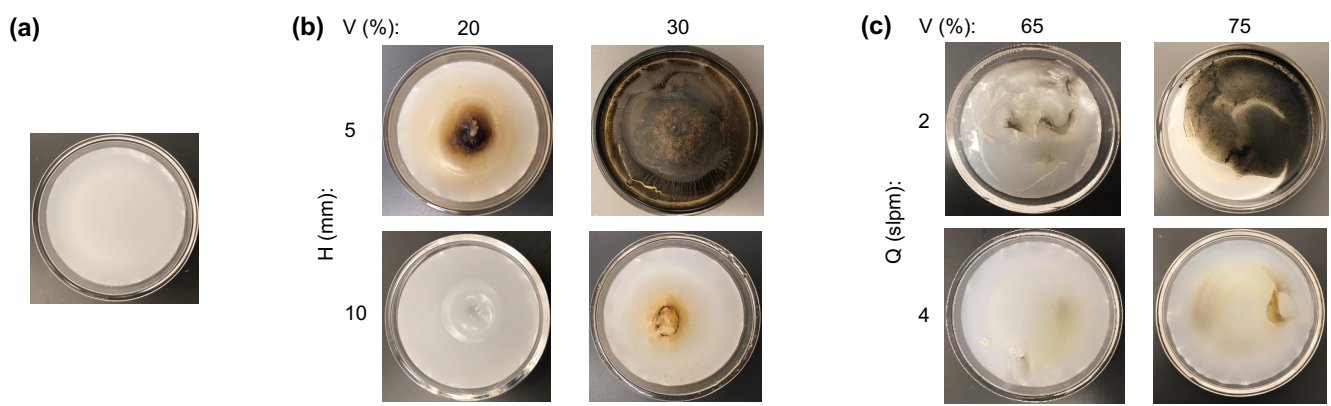

**Fig. 9 - LDPE sample treatment. (a) Pristine sample before plasma treatment and samples after 30 minutes of plasma treatment under representative operating conditions in the (b) transarc and (c) glidarc reactors.**

**Figure 9c** shows the samples after 30 minutes of treatment in the glidarc reactor. The samples melt within five minutes of the experiment under all the selected operational conditions. The darkening of the sample's surface is more significant for the lower flow rate ($Q$ = 2 slpm) and higher voltage level ($V$ = 75%), consistent with the enhanced interactions between the plasma and the feedstock revealed by Schlieren imaging (section 3.4). Higher voltage levels lead to greater plasma power deposited on the feedstock, as suggested by the higher heat fluxes in the simulation results in **Fig. 3d**. The lower flow rate of 2 slpm leads to a longer interaction time between the reactive plasma species and the feedstock (which can be assumed proportional to $t_{res}$). In contrast, the higher flow rate of 4 slpm, which leads to a shorter $t_{res}$ of 10.3 s and greater convective cooling of the plasma, results in minor darkening of the feedstock, as observed in **Fig. 9c**. This is also supported by the simulation results shown in section 2.2, in which the heat flux and temperature in the glidarc reactor are higher for $Q$ = 2 slpm as compared to $Q$ = 4 slpm due to lower convective cooling. As discussed in section 4.3, the lower flow rate and higher voltage level used in the glidarc reactor lead to more significant plasma-feedstock interaction, which favors greater hydrogen production.

### 4.2 Electrical characterization

The electrical characterization of the reactors helps assess the dynamics of the plasma and determine their role in hydrogen production. The transarc plasma produces a smooth sinusoidal voltage signal of up to 22.5 kV peak-to-peak (pp) with a sharply varying current of frequency ~ 25 kHz. The sharply varying current is characteristic of filamentary (spark) discharges. The glidarc plasma generates smooth sinusoidal signals of frequency ~ 22.5 kHz and an instantaneous voltage of up to 6 kV pp, which is in phase with the current.

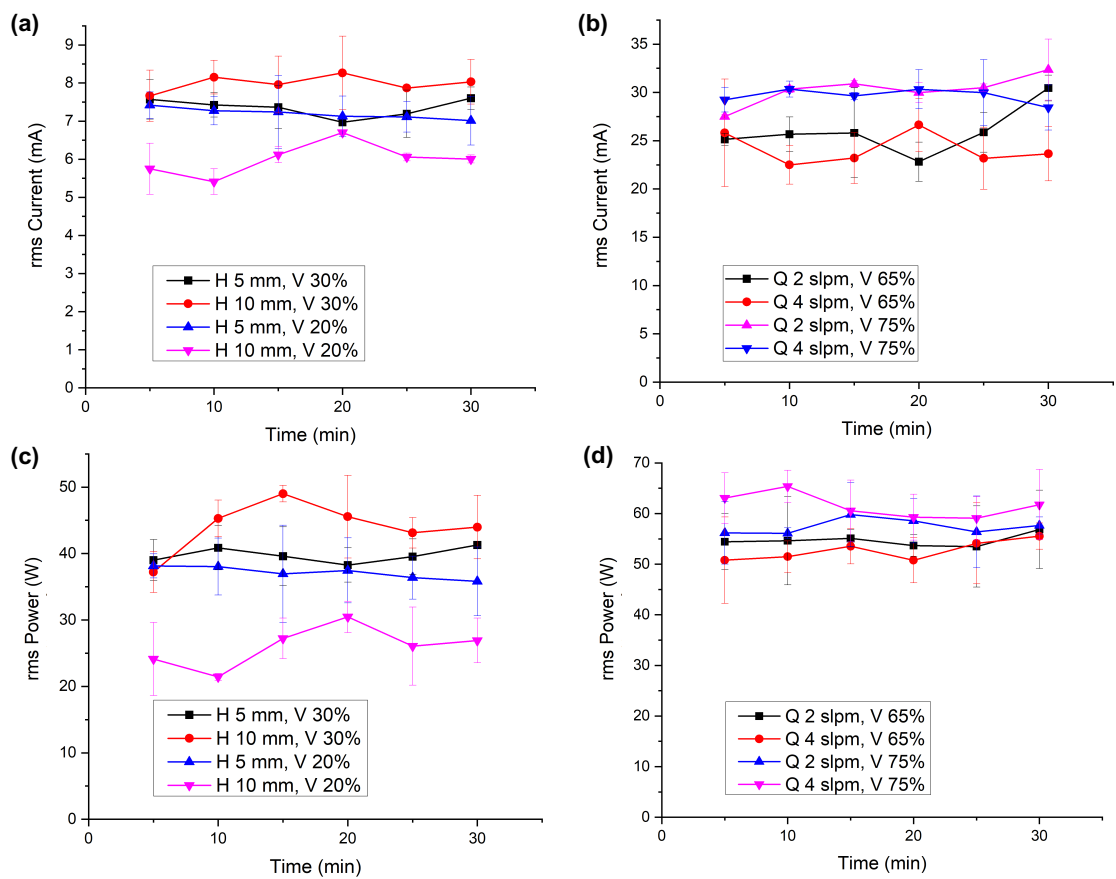

**Fig. 10 - Electrical characteristics of reactors operation. Root-mean-square (rms) current as a function of time for the (a) transarc and (b) glidarc reactors and rms power for the (c) transarc and (d) glidarc reactors.**

The overall electrical characteristics as a function of operating conditions are presented in **Fig. 10**. The root-mean-square (rms) current increases with increasing voltage level for both reactors (**Fig. 10a** and **Fig. 10b**), resulting in increased power deposited over the feedstock. Also, for the transarc, a higher rms current implies higher electron flux onto the feedstock, which can likely increase the probability of cleavage of carbon-hydrogen (C-H) bonds and consequently increase hydrogen production. In comparing the results in **Fig. 10a** with those in Fig. **10b**, it is to be noted that the glidarc reactor utilizes two power supplies and hence uses significantly greater current than the transarc reactor. The rms power as a function of time for both reactors and the different experimental conditions are presented in **Fig. 10c** and **Fig. 10d**. In both reactors, voltage level $V$ is the main parameter determining the consumed power, which increases with $V$ irrespective of other conditions (i.e., electrode-feedstock spacing $H$ and flow rate $Q$). The transarc plasma rms power (**Fig. 10b**) increases slightly with voltage level and fluctuates minimally during the experiments for all the operational conditions tested. Similarly, as for the glidarc reactor, the rms power for the glidarc reactor increases slightly with voltage level. The maximum rms power is obtained for $Q = 4$ slpm and $V = 75\%$, as shown in **Fig. 10d**. Additionally, the rms power of the glidarc reactor shows negligible variation across the different conditions tested, and it is approximately 20 W higher than that of the transarc. Despite the glidarc reactor's sensitivity to flow rate, the effect of flow rate on rms power is negligible. However, the residence time $t_{res}$ is shorter at higher flow rates, limiting the interaction time between plasma species and the feedstock, potentially lowering hydrogen production.

**4.3 Process performance**

The main performance metrics of the process are the hydrogen production rate and the hydrogen production efficiency (i.e., hydrogen production rate per unit power). The hydrogen production rate ($P_r$) is defined as:

$$P_r = C_{out}Q, \qquad (2)$$

where $C_{out}$ is the molar concentration of hydrogen in the outflow stream. The hydrogen production efficiency ($\eta_e$) is given by:

$$\eta_e = \frac{P_r}{P_{rms}}, \qquad (3)$$

where $P_{rms}$ is the rms power consumed by the reactor.

Results for hydrogen production rate $P_r$ as a function of process time and operational parameters for both reactors are shown in **Fig. 11**. The results show that $P_r$ increases with increasing voltage level in both reactors. The transarc reactor (**Fig. 11a**) attains the mean maximum production rate of 6.6 mmol/h (0.33 mmol/g LDPE) at $H = 5$ mm, $Q = 0.1$ slpm, and $V = 30\%$; while the maximum average production rate for the glidarc reactor (**Fig. 11b**) is 8.4 mmol/h (0.42 mmol/g LDPE) at $Q = 2$ slpm, $H = 5$ mm, and $V = 75\%$. These conditions for maximum hydrogen production correspond to those observed by Schlieren visualization (section 3.4), namely, greatest turbulence for the transarc and greatest plasma-substrate interaction for the glidarc, respectively. In general, greater hydrogen production is attributed to the larger amount of power deposited over the LDPE sample at higher voltage levels. Furthermore, $P_r$ increases with time under higher voltage levels during the first ~15 minutes and then stabilizes. The slight decline in hydrogen production rate after 25 minutes likely suggests the formation of a layer of carbon/char that hinders hydrogen production. The lower hydrogen production rates at the beginning of the experiments (< 10 minutes) under all operational conditions and in both reactors suggest that a portion of the energy is consumed in melting the samples. After melting, the energy deposited by the plasma may lead to a more effective incision of C-H bonds in LDPE, leading to greater $P_r$.

Despite lower voltage level, the hydrogen production rate for $H = 5$ mm, $Q = 0.1$ slpm, and $V = 20\%$ is higher than that of $H = 10$ mm, $Q = 0.1$ slpm, and $V = 30\%$ (lines with triangle and circle marks, respectively, in **Fig. 11a**). This suggests that shorter electrode-feedstock spacing $H$ leads to greater interaction between the reactive plasma species and the LPDE feedstock, resulting in higher hydrogen production rates. On the contrary, when $H$ is larger, the highly energetic electrons and reactive plasma species lose significant energy due to collisions resulting in quenching or recombination reactions before having the opportunity to interact with the feedstock, leading to lower hydrogen production rates.

The results in **Fig. 11b** show that, for the glidarc reactor, greater flow rates lead to lower hydrogen production rates. This is likely due to two effects. First, the shorter residence time $t_{res}$ reduces the probabilities of electrons and excited species reacting with the feedstock. This has been observed by Indarto *et al*. [47] in investigating the effect of working gas flow rate on carbon dioxide conversion using glidarc plasma. Second, intense convective cooling of the glidarc plasma leads to lower heat flux over the feedstock, as depicted in the simulation results in section 2.2. The higher hydrogen production rate for $Q = 2$ slpm, $V = 75\%$, and $H = 5$ mm is attributed to longer $t_{res}$, less cooling, and higher deposited power. The slightly higher hydrogen production rate by the glidarc reactor is due to its utilization of two power sources, which effectively increases the electrical power deposited on the feedstock.

To determine how effectively electrical energy is utilized to produce hydrogen from LDPE, hydrogen production efficiency $\eta_e$ is shown in **Fig. 11c** and **Fig. 11d** for the transarc and glidarc reactors, respectively. Hydrogen production efficiency increases proportionately with voltage levels for both reactors. High voltage levels lead to greater power deposited onto the feedstock, which increases hydrogen production rate irrespective of the other operating conditions. This observation is consistent with the computational simulation results of greater heat flux onto the feedstock (section 2.2). Additionally, as observed in the hydrogen production results in **Fig. 11a** and **Fig. 11b**, electrode-feedstock spacing and flow rate significantly affect the hydrogen production efficiency in the transarc and the glidarc reactor, respectively. The transarc reactor attains a maximum $\eta_e$ of 0.16 mol/kWh at $H = 5$ mm, $V = 30\%$, and $Q = 0.1$ slpm. This maximum $\eta_e$ is comparable to that of the glidarc reactor, which is 0.15 mol/kWh at $Q = 2$ slpm, $V = 75\%$, and $H = 5$ mm. Overall, despite markedly different modes of operations, hydrogen production rate and hydrogen production efficiency are similar in both reactors for all operational conditions. This is an important aspect to factor in for the scaling-up of the systems.

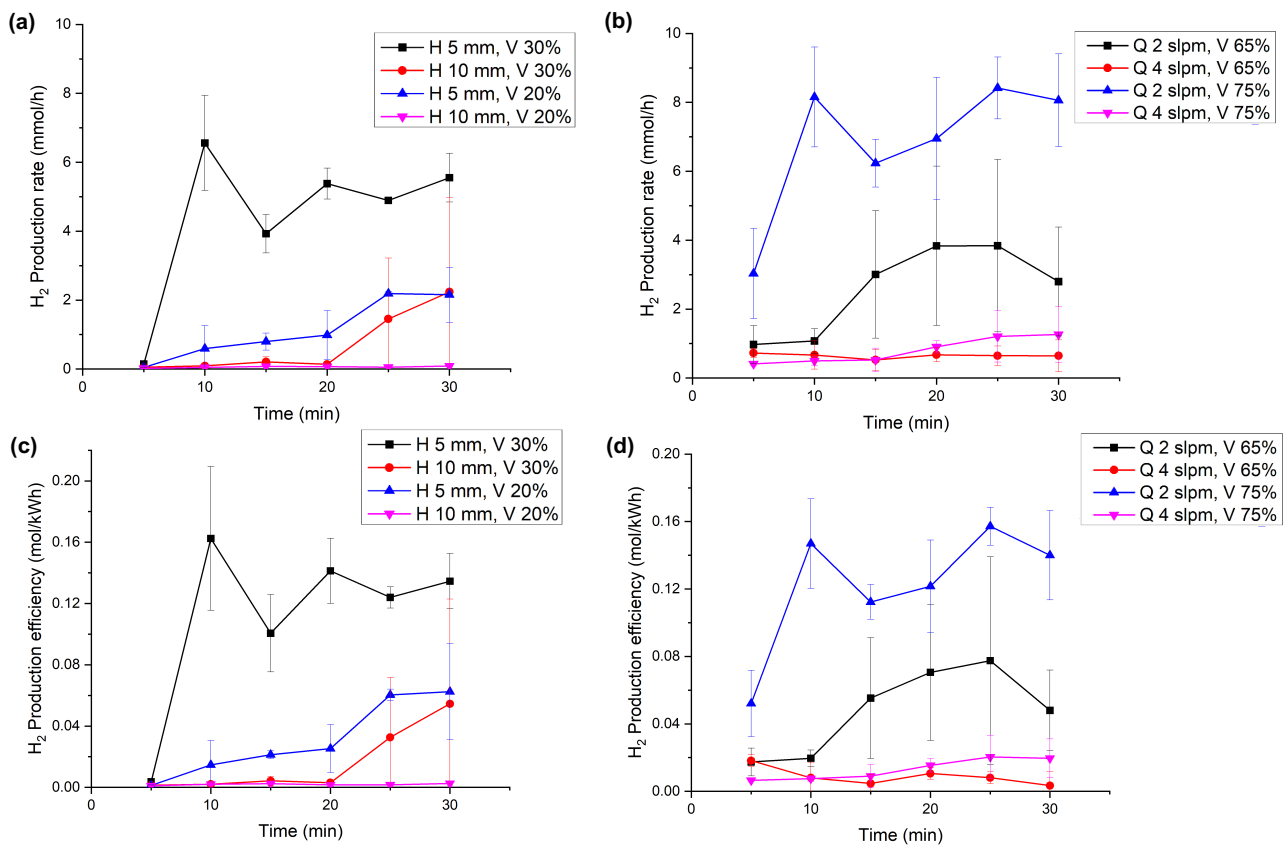

**Fig. 11 - Hydrogen production rate versus time for the (a) transarc reactor and the (b) glidarc reactor. Hydrogen production efficiency as a function of time for the (c) transarc reactor and (d) the glidarc reactor.**

To ascertain the economic viability of hydrogen production using the two reactors, the energy cost of hydrogen production, defined as the amount of energy required for production of 1 kg of hydrogen [48] from LDPE, is evaluated and compared with steam methane reforming and water electrolysis, the most efficient processes. The energy cost of hydrogen production for transarc and glidarc reactors are 3100 and 3300 kWh/kg $H_2$, respectively, and are significantly higher than that of steam methane reforming (21.9 kWh/kg $H_2$) and electrolysis (47.6 kWh/kg $H_2$). Perhaps, it is ascribed to the nature of the feedstock; methane and water vapors feedstocks (both gaseous) require less energy in overcoming the weak intermolecular forces as well as cleaving atomic bond energies as compared to a solid LDPE with stronger intermolecular forces and atomic bonds. Although hydrogen production from LDPE via low-temperature atmospheric plasma has a higher energy cost of hydrogen production, the process potentially has greater environmental benefits, particularly when powered by renewable energy sources such as wind and solar energy

### 4.3 Correlations between operational parameters and hydrogen production

The expected performance of the reactors, necessary for scaling analyses, can be assessed through correlations between hydrogen production ($P_r$ and $\eta_e$) and operational parameters ($V$, $Q$, $H$) and/or operational characteristics (e.g., rms voltage $V_{rms}$, rms power $P_{rms}$). Correlations of the form $V_{rms}^a P_{rms}^b H^c$ and $V_{rms}^a P_{rms}^b Q^c$ are sought for the transarc and the glidarc reactors, respectively. For dimensional and practical reasons, the exponents were set as $a, b, c \in \{-3, -2, -1, 0, 1, 2, 3\}$. This set of exponents leads to 343 ($7^3$) different parametric combinations. Among these, the conditions with the strongest correlation are identified as those with the greatest correlation coefficient ($R^2$), which are depicted in the results in **Fig. 12**.

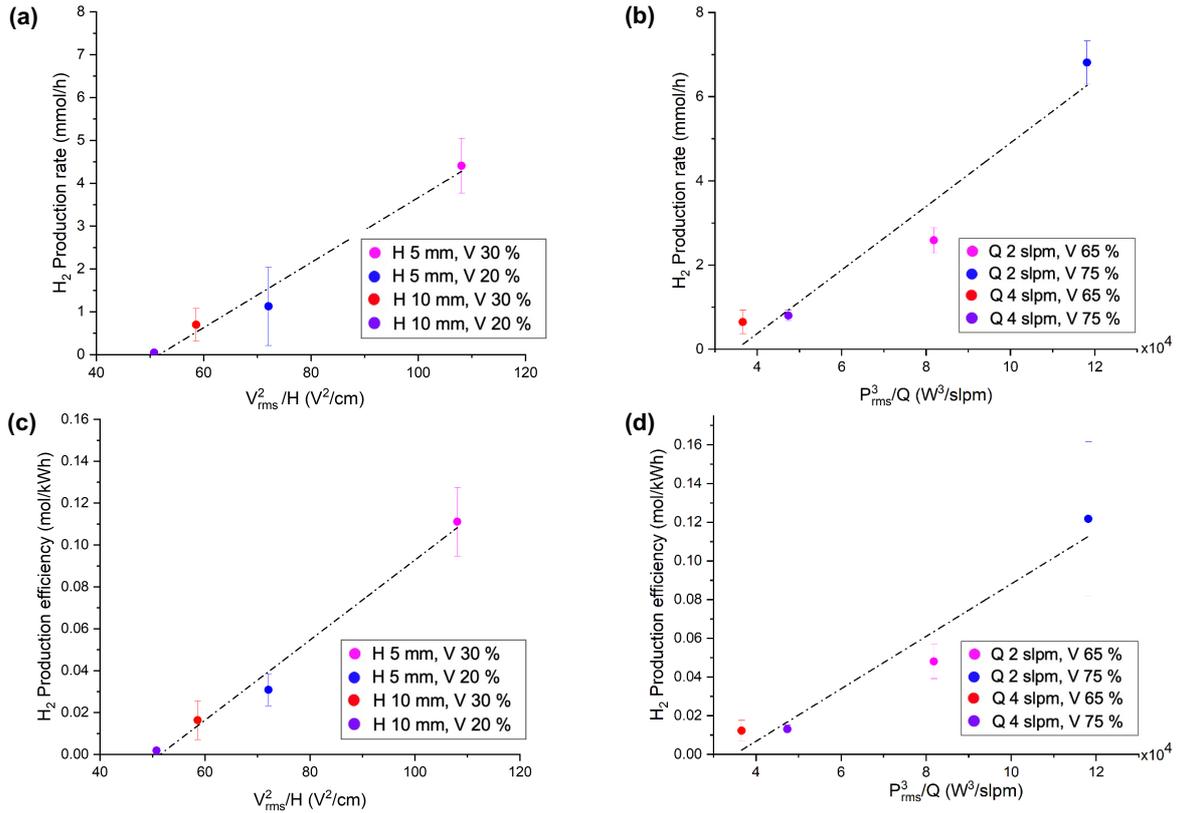

**Fig. 12 – Correlations between hydrogen production and operating parameters. Hydrogen production as a function of (a) equivalent power per unit length for the transarc reactor and (b) power per unit flow rate for the glidarc reactor. Hydrogen production efficiency as a function of (c) equivalent power per unit length for the transarc reactor and (d) power per unit flow rate for the glidarc reactor.**

The hydrogen production rate $P_r$ by the transarc reactor strongly depends on the rms voltage and electrode-feedstock spacing $H$ following the relation: $P_r = \alpha_r V_{rms}^2/H - \beta_r$ ($\alpha_r = 7.582 \times 10^{-4}$, $\beta_r = 3.918$, $R^2 = 0.979$), which is shown in **Fig. 12a**. Given the columnar structure of the transarc plasma, the term $V_{rms}^2/H$ can be considered equivalent to plasma power per unit length. Hence, the obtained correlation implies that $P_r$ for the transarc reactor is correlated with plasma power unit length. This result is consistent with the simulation results in section 2.2 indicating a direct dependency of surface heat flux and of average temperature with thermal power per unit volume. The inverse dependency of $P_r$ with electrode-feedstock spacing $H$ is also consistent with the findings in section 4.3. Similarly, hydrogen production efficiency strongly correlates with plasma power per unit length, leading to a linear relationship given by $\eta_e = \alpha_e V_{rms}^2/H - \beta_e$ ($\alpha_e$=1.911×10$^{-5}$, $\beta_e$ = 9.828×10$^{-2}$, R$^2$ = 0.986). For a given plasma power (proportional to $V_{rms}^2$), $\eta_e$ increases monotonically with decreasing inter-electrode feedstock spacing $H$.

Both, hydrogen production rate $P_r$ and production efficiency $\eta_e$ for the glidarc reactor depend on the cube of the rms power $P_{rms}$ and inversely to the flow rate $Q$, as shown in **Fig. 12b** and **Fig. 12d**. The inverse relationship between $P_r$ and $Q$ is consistent with the simulation results in section 2.2 in which higher temperature and surface heat flux are observed for the lower flow rate of 2 slpm. A low flow rate produces less convective cooling and higher residence time $t_{res}$, which lead to longer characteristic times for plasma species to interact with the feedstock. The correlation of $P_r$ with $P_{rms}$ and $Q$ is given by $P_r = \alpha_r P_{rms}^3/Q - \beta_r$ ($\alpha_r = 7.546 \times 10^{-5}$, $\beta_r = 2.646$, R$^2$ = 0.940). It is to be noted that in the absence of inflow gas ($Q$ = 0 slpm), the generated plasma does not glide down along the electrodes, and hence does not interact with the feedstock (leading to no hydrogen production). Contrastingly, for the larger flow rates, the hydrogen production rate reduces significantly due to rapid cooling of the gas and limited time for plasma species to interact with the feedstock. Therefore, optimal hydrogen production is attained at intermediate values of $Q$, as depicted in the results in **Fig. 6**. The dependence of the production rate with the cube of rms power suggests a trend that compensates for the significant amount of energy consumed at the beginning of the sample treatment (i.e., slow melting of the top of the sample compared to what is achieved by the transarc), and then the production increases (**Fig. 11b**). The production efficiency $\eta_e$ of the glidarc reactor depicts a comparable trend as $P_r$ given by the relation $\eta_e = \alpha_e P_{rms}^3/Q - \beta_e$ ($\alpha_e$=1.355×10$^{-6}$, $\beta_e$ = 4.74×10$^{-2}$, R$^2$ = 0.945). The greater residence for lower flow rates implies greater plasma interaction with the feedstock leading to greater production efficiency. The hydrogen production efficiency's dependency on the flow rate is limited to a specified range as no plasma-feedstock interaction is achieved at too low ($Q$ < 2 slpm) or high ($Q$ > 6 slpm) flow rates (see section 3.2).

## 5. Conclusions

Two nonthermal plasma reactors with complementary characteristics, based on transarc and glidarc discharges, are designed, developed, and characterized to produce hydrogen from LDPE as a model plastic waste. CFD thermal-fluid models are used to attain expected operational characteristics as functions of design and operation parameters, namely electrode-feedstock spacing, flow rate, and dissipated thermal power – the latter assumed proportional to the voltage level of the power supply. Simulation results identify electrode-feedstock spacing, flow rate, and voltage level as the main process parameters of the reactors. The built reactors are experimentally evaluated using electrical diagnostics, optical and Schlieren imaging, and gas chromatography to quantify hydrogen production. The Schlieren visualization results qualitatively show that hydrogen production correlates with the amount of turbulence over the LDPE feedstock for the transarc reactor and the residence time of the plasma over the feedstock for the glidarc reactor. The experimental evaluation of hydrogen production from LDPE shows that the power consumed by the plasma remains approximately constant throughout the 30 min treatment time. Moreover, the results show that hydrogen production increases proportionally with voltage level in both reactors and that electrode-feedstock spacing and flow rate are the dominant operational parameters in the transarc and the glidarc reactor, respectively. The energy cost of hydrogen production for both reactors is significantly higher than the conventional and most efficient hydrogen production technologies of steam methane reforming and water electrolysis. Hydrogen production and production efficiency correlate linearly with rms voltage squared divided by inter-electrode spacing for the transarc reactor and with the rms power cubed divided by flow rate for the glidarc reactor. The two reactors depict comparable performance in terms of hydrogen production rate and efficiency, despite distinct differences in their operational principle. Overall, the results show that atmospheric pressure nonthermal plasma is effective at producing hydrogen from LDPE.

**Acknowledgments**

This work has been supported by the US Army Combat Capabilities Development Command (CCDC) Soldier Center Contracting Division through Contract # W911QY-20-2-0005.